\documentclass[aps,twocolumn,showpacs,english]{revtex4}

\usepackage[T1]{fontenc}
\usepackage[latin9]{inputenc}
\usepackage{textcomp}
\usepackage{amsmath}
\usepackage{graphicx}
\usepackage{amssymb}
\usepackage{babel}

\usepackage{natbib}

\begin{document}
\bibliographystyle{plainnat}

\title{Exciton-phonon coupling efficiency in CdSe quantum dots \\ embedded in ZnSe nanowires}

\author{S. Bounouar$^{1,2}$, C. Morchutt$^{2}$, M. Elouneg-Jamroz$^{1,2}$, L. Besombes$^{1}$, R. Andr\'{e}$^{1}$, E. Bellet-Amalric$^{2}$, C. Bougerol$^{1}$, M. Den Hertog$^{3}$, K. Kheng$^{2}$, S. Tatarenko$^{1}$,  and J. Ph.~Poizat$^{1}$}

\affiliation{
$^1$ CEA-CNRS-UJF group 'Nanophysique et Semiconducteurs', Institut N\'{e}el, CNRS - Universit\'{e} Joseph Fourier, 38042 Grenoble, France, \\
$^2$ CEA-CNRS-UJF group 'Nanophysique et Semiconducteurs', CEA/INAC/SP2M, 38054 Grenoble, France, \\
$^3$ Institut N\'{e}el, CNRS - Universit\'{e} Joseph Fourier, 38042 Grenoble, France,}

\begin{abstract}

Exciton luminescence of a CdSe quantum dot (QD) inserted in a ZnSe nanowire  is
strongly influenced by the dark exciton states. Because of the small size of
these QDs (2-5nm), exchange interaction between hole and electron is highly
enhanced and we measured large energy splitting between bright  and dark exciton
states ($\Delta E\in [4, 9.2 ]$ meV) and large spin flip rates between these  states.
Statistics on many QDs showed that this splitting depends on
the QD size. Moreover, we measured an increase of the spin flip rate to the
dark states with increasing energy splitting. We explain this observation with a
model taking into account the fact that the exciton-phonon interaction depends
on the bright to dark exciton energy splitting as well as on the size and shape
of the exciton wave function. It also has consequences on the exciton line intensity
at high temperature.

\end{abstract}


\maketitle

\section{Introduction}

Semiconductor nanowires (NWs) have attracted great attention in the last few years since they hold great promise to become building blocks in tomorrow's nanoscale devices and circuits with  vast potential applications ranging from nanoelectronics \citep{Duan01,Lu,The03}, optoelectronics (light emitting diodes  \citep{Konenkamp,Kim}, nanolasers \citep{Duan03}),  thermoelectrical energy conversion  \citep{Hochbaum}, to biological or chemical sensors \citep{Cui}.
Most of the NW growth methods allow for the variation of the chemical composition \citep{Gud02,Bjork} or doping \citep{Yang} along the longitudinal or radial directions. This enables the fabrication of semiconductor heterostructures, and more specifically of  quantum dots (QDs).

Single  QDs have turned out to be excellent candidates for stable and efficient single photon sources \citep{singleemitterimamoglu,Moreau,Santori,Chang,Claudon}. Within this category, QDs embedded in NWs have already demonstrated single photon emission \citep{borgstrom}, even at relatively high temperature ($T=220$K) \citep{TRIBU220k}. Quest for efficient and eventually room temperature QD single photon sources requires a good understanding of the excitonic dynamics in such systems.

 In a QD,  exciton states are split by the electron hole exchange
   interaction into  higher energy bright exciton states (BS) and  lower
   energy dark states (DS) \citep{dark} with respective angular momentum of
   $J_z=\pm1$ and $J_z=\pm
   2$. In the present system, the energy splitting is relatively
   large, and has already been measured around $\Delta E=6$ meV \citep{sallen}.
   As exchange interaction in semi-conductor materials is proportional to the
   spatial overlap between the electron and hole wave functions, it is strongly
   enhanced in low dimensional objects. Values of $\Delta E \in [2, 4] meV$ have been measured in
   quantum wells \citep{quantumwelldarkexc} and it has been demonstrated that they can be even higher in quantum wires \citep{quantumwire}.
   Extremely large splitting were calculated and observed \citep{efrosexp} in colloidal QDs ($\Delta E\in[2, 20]$ meV).
   This is the result of very good confinement of the carriers owing to the  small
   size of the QDs.
Effects of the dark states on the QD luminescence become noticeable when $\Delta E \gg k_B T$, with $k_B$ the Boltzmann
constant and $T$ the temperature \citep{sallen,lounistempdep,influencedsonqd}.
   We show that under non resonant pumping, the BS excitonic population leaks towards the DS that recombines non-radiatively, leading to a reduced excitonic light emission compared to the biexcitonic one.   Transitions between the BS and DS states are due to  hole spin flips assisted by phonons \citep{holespinflip}.
   The speed of this processes depends essentially on the efficiency of the hole interaction with the phonon reservoir experienced by the QD \citep{phonon-exciton_spectraldensity}.

   In this paper, we show that  the QD size
   can highly influence the QD phonon spectral density  and modify
   the spin flip rates between the bright and dark exciton states.
   It not only has consequences on the exciton intensity at $T=4$K but also at
   high temperatures.

The paper is organized as follows. The sample  preparation and the setup are presented in section \ref{sample}. In section \ref{DeltaE} the BS to DS energy splitting $\Delta E$ of several QDs is extracted using temperature dependent lifetime measurements. In section \ref{phonon} we present the experimental dependence of the spin flip rate versus  $\Delta E$, and suggest an explanation based on a theoretical model. These results are then used in section \ref{exciton} to discuss the exciton to biexciton line intensity ratio as a function of $\Delta E$ and of the temperature.


\section{Sample and Setup}
\label{sample}

The nanowires are grown by molecular beam epitaxy using the vapor-liquid-solid
technique \citep{Hertog}. The substrate is GaAs 100 with a ZnSe buffer. After
dewetting of a thin layer of gold at 500°C, the growth is performed at 410°C.
The nanowires have a diameter of 10 nm and a length of 400 nm. The wire
diameter (around 10 nm) is of the order of the bulk exciton Bohr diameter in
CdSe (11nm), which means that the carriers in the QD are in the strong
confinement regime. In order to study single QD, the nanowires are broken and
dispersed on a silicon substrate by direct contact. High resolution transmission electron microscope (TEM) images
revealed very small QD sizes, from 2 to 5 nm in the nanowire
direction \citep{JAP}. Other TEM experiments coupled to spectroscopy on other samples
showed that this growth technique often led to ZnSe encapsulated QDs.
Photoluminescence on
as-grown samples is centered around 2,35eV with a dispersion of $\pm 0.07eV$.

The experimental apparatus is a time resolved microphotoluminescence experiment
setup. The excitation source is a frequency doubled picosecond Ti sapphire
laser emitting at a wavelength of $\lambda =950/2=475$ nm (ie $2.6$ eV).
Nanowires luminescence is detected through a $\delta \lambda=0.04$ nm (ie
$\delta E=0.2$ meV) resolution spectrometer on a charged coupled device (CCD)
camera for spectra or on a low jitter (40ps) avalanche photodiode (APD) for time resolved measurements (60 ps time resolution for the whole set-up).

\section{QD Size effect on the exchange interaction}
\label{DeltaE}

\begin{figure}
\resizebox{0.45\textwidth}{!}{\includegraphics[width=0.4\textwidth]{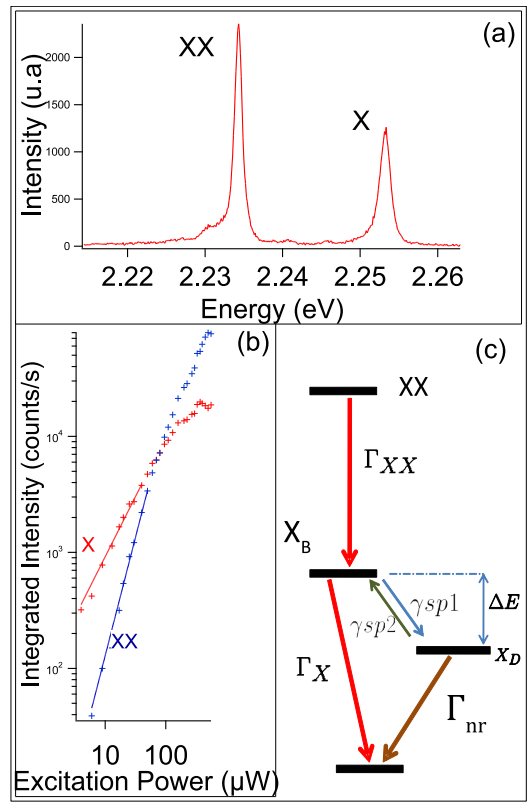}}
\caption{(Color online) (a) Spectrum of a neutral QD at $T=4$ K. (b) Power dependence of the X and XX lines under pulsed
excitation at $T=4$ K. (c) Level scheme and transition rates of a neutral QD.} \label{fig:spectre}
\end{figure}

Figure \ref{fig:spectre}(a) shows typical spectrum obtained on a single
nanowire. The two transitions correspond to exciton and biexciton (respectively
denoted X and XX) and were identified using cross-correlation techniques
\citep{sallen}. The biexciton binding energy is measured around 20 meV. The four
relevant states of a neutral QD are represented on figure
\ref{fig:spectre}(c). The exciton is splitted in two states (noted DS and BS)
linked by spin flip transition rates noted $\gamma_{sp1}$ and $\gamma_{sp2}$.
The quantities $\Gamma_X$ and $\Gamma_{XX}$ are the radiative recombination
rates of the bright exciton and the biexciton, and $\Gamma_{NR}$ is the non
radiative decay rate of the dark exciton. These quantities have been measured at $T=4$ K
on nine QDs
as $1/\Gamma_X= 0.50  \pm 0.05$ ns, $1/\Gamma_{XX}= 0.30  \pm 0.05$ ns, and
$1/\Gamma_{nr}= 5.0  \pm 0.5$ ns, where the error is the dispersion amongst QDs.
An example of power dependence under pulsed excitation is shown in figure
\ref{fig:spectre}(b) evidencing the linear and quadratic behavior of the X and XX line intensity respectively. In all the investigated QDs, saturating intensity of the
exciton emission is smaller than the biexciton, which is a signature of the
strong influence of the dark exciton \citep{influencedsonqd}. When $\Delta E\gg
k_BT$, dark excitons cannot transit back to the bright state as there are no
phonons available for such a spin flip and dark excitons are stored until
recombining non radiatively. As a result, the photoluminescence  of the exciton is less intense
than that of the biexciton at saturation. The biexciton luminescence is not
affected by the presence of the DS, whereas the exciton BS has a large
probability to decay to the DS and not produce any photon. As spin flip rates
are here of same order of magnitude or larger than the BS radiative
decay, exciton intensity is generally small compared to the biexciton.

\begin{figure}
\resizebox{0.4\textwidth}{!}{\includegraphics*[0,150][720,580]{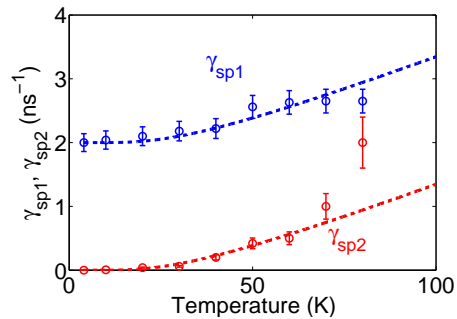}}
\caption{(Color online) Transition rates $\gamma_{sp1}$ and $\gamma_{sp2}$ versus temperature.  The dashed lines are the plot from the theoretical model following phonon population evolution with
temperature.} \label{fig:split}
\end{figure}

 Temperature dependence of exciton decay
time  give access to the energy splitting $\Delta E$ between DS and BS
\citep{lounistempdep, sallen}. Spin flip transition rates depend on the spectral
density of the acoustic phonons at the energy of the transition $\Delta E$ and
on the phonon population that follows a Bose-Einstein distribution $N_B(\Delta
E,T)=1/(1-\exp(\Delta E/k_B T)) $. The spin flip rates
$\gamma_{sp1}$ (bright to dark) and $\gamma_{sp2}$ (dark to bright) can be
written as:
\begin{eqnarray}
 \gamma_{sp1} & \propto &
(N_B(\Delta E,T)+1)R(\Delta E), \\
    \gamma_{sp2} & \propto & N_B(\Delta E,T)R(\Delta E) .
    \end{eqnarray}
The quantity $R(\Delta E)$ depends on the BS-DS energy splitting $\Delta E$ but
also on the spatial extension and shape of the wave function of the
exciton\citep{lucienphononbroad,Takagahra}.

At $T= 0$K,  $N_B=0$, so that the spectral density $R(\Delta E)\propto \gamma_{sp1}(0K)$ represents the bright
to dark state transition rate and $\gamma_{sp2}=0$. When the temperature is
increased so that $k_BT\sim \Delta E$, the bright exciton state is repopulated.
For each temperature the transition rates $\gamma_{sp1}$ and $\gamma_{sp2}$ are
extracted from the time resolved luminescence and their evolution is fitted
using the model based on the level scheme of figure \ref{fig:spectre} (c)) with $\Delta E$  as the fitting parameter.

\begin{figure}
\resizebox{0.45\textwidth}{!}
{\includegraphics*[0,200][720,660]{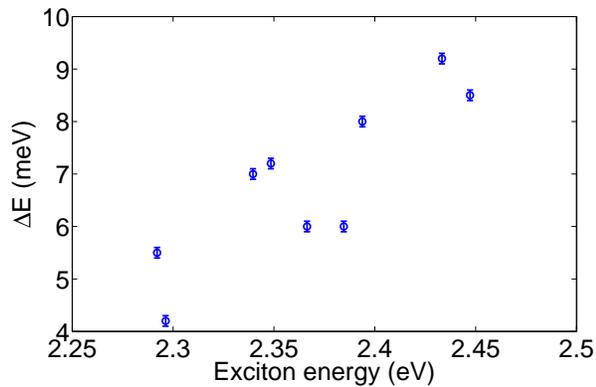}}
\caption{(Color online) Increase of splitting energies $\Delta E$ between DS and BS measured on nine QDs
as a function of their excitonic emission energy.  } \label{fig3:excposvssplit}
\end{figure}

This process is repeated on nine QDs. As shown on figure \ref{fig3:excposvssplit},
measured splittings range from 4.2 meV to 9.2 meV. A clear trend is appearing:
splittings measured for low energy emission QDs are generally smaller compared
to splittings measured for high energy emission QDs. The latters correspond to small size QDs, where electron
and hole wave functions  overlap very well, so that their exchange
energy is large, leading to large $\Delta E$. Such relationships between QD sizes and
energy splittings have been calculated  for colloidal QDs
\citep{efros}. Change in composition of the Cd$_x$Zn$_{1-x}$Se QD can also affect
the emission energy and the value of $\Delta E$. Our measurements do not follow
exactly a smooth law.   The shape of the confinement has also a strong
influence on the wave function forms, and consequently on their correlation
function. For example the prolate or oblate nature of the QD  geometry
appears to have an important effect. This explains why we obtained a cloud of
experimental points following a general trend instead of a strict dependence.

\section{Efficiency of the exciton-phonon coupling}
\label{phonon}

\begin{figure}
\resizebox{0.5\textwidth}{!}
{\includegraphics*[0,200][720,650]{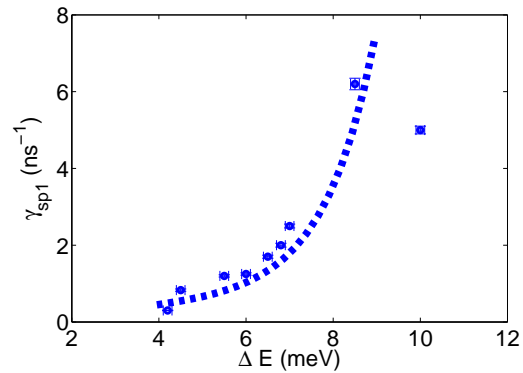}}
\caption{(Color online) Transition rate from BS to DS $\gamma_{sp1}$ at 4K plotted as a function of
splitting energy $\Delta E$ between DS and BS for nine differents QDs. The dotted line is
a guide for the eyes.} \label{fig:gammaovssplit}
\end{figure}

In figure \ref{fig:gammaovssplit}, we have plotted the transition rates from BS to DS
measured at 4K, $\gamma_{sp1}(4K)$, versus energy splittings $\Delta E$ for nine different QDs.
A clear enhancement of the transition rate for large $\Delta E$ can be observed.
It can be also noted that the measured rates are comparable and even large than the excitonic radiative rate   \citep{influencedsonqd}.
 This is why the
bright exciton is so depopulated compared to the biexciton in all the QDs investigated.

 The results displayed in figure \ref{fig:gammaovssplit} indicate that large $\Delta E$ splittings lead to extremely fast depopulation of the bright
state suggesting that exciton-phonon interaction becomes more efficient with
large $\Delta E$. As explained, larger splittings  correspond to
small QDs. Exciton coupling to phonons not only depends on the energy needed
for the spin flip but also on the QD size. But size and $\Delta E$ are not independent parameters and induce opposite effects on the transition rates. In order to explain this non-trivial
behavior, it is necessary to calculate the phonon spectral density as
a function of QD size and energy splitting $\Delta E$ for each particular QD to obtain a general trend.

We shall only consider longitudinal acoustical (LA) compression modes since longitudinal optical (LO) phonons have energies of about $30$ meV far above the dark - bright exciton splitting $\Delta E$. In the following we will perform  this calculation both in the 3D  and in the 1D case for the available phonon modes.
The NW geometry is obviously somewhere in-between these two extreme cases that will only give us a qualitative behavior.

 We first consider the 3D situation.
The phonons dispersion is bulk-like, approximated by the Debye law
 $w(k)=c_lk$, with $c_l$ the sound speed for LA phonons in the semiconductor material.
Piezoelectric interactions and Fr\"{o}hlich longitudinal optical (LO) phonon couplings are neglected.
The exciton-phonon interaction is dominated by the hole-phonon interaction \citep{holespinflip}. We therefore consider only
 the latter, whose Hamiltonian  can be written, in
the second quantization representation with respect to the carrier states, as:
\begin{equation}
  \label{}
  H_{h} = \frac{1}{\sqrt N}\sum_{knn'}a_n^\dag a_n'f_{h,nn'}(k)(b_{k}+b_{-k}),
\end{equation}
where $a_n^\dag$ and $a_n$ are the hole creation and annihilation
operators, $b_{k}^\dag$ and $b_{k}$ are LA phonon creation and annihilation
operators. The index $n$ represents the excitation level of the hole, and $k$ the phonon mode.
The coupling constant is defined as
\begin{equation}
  \label{}
 f_{h,nn'}(k)= \sigma_{h} \sqrt{ \frac{\hbar k}{2 \rho Vc_l}}F_{nn'}(k),
\end{equation}
where $\sigma_{h}$ is the deformation potential for  holes, V is the
unit cell volume, $\rho$ is the CdSe volumic mass.
The quantity $F_{nn'}(k)$ is a purely geometrical form factor given by
\citep{phonon-exciton_spectraldensity} :\begin{equation}
  \label{}
 F_{nn'}(k)= \int_{-\infty}^{\infty}d^3r\psi_n^*(r)e^{ikr}\psi_{n'}(r).
\end{equation}

In order to evaluate the coupling constant for the lowest hole state
$f_{11}(E)$ we consider the harmonic oscillator potential ground state as the
wave function of the hole:
\begin{equation}
\psi(r)=\frac{1}{\pi^{3/4}l_\bot\sqrt{l_z}}\exp \left[-\frac{1}{2}\left(\frac{r{_\bot}}{l_{\bot}}\right)^2-\frac{1}{2}\left(\frac{z}{l_z}\right)^2 \right],
\end{equation}
where $r_\bot$ is the position component in the xy plane and $l_\bot,l_z$ are
respectively the in plane, and out of plane ($z$ direction)
localization widths. For this
wave function the form factor is easily found as
\begin{equation}
F_{11}(k)=\exp\left[-\left(\frac{k{_\bot}l_{\bot}}{2}\right)^2-\left(\frac{k_zl_z}{2}\right)^2\right].
\end{equation}
For the lowest hole state, the phonon spectral density in the QD is:
\begin{equation}
\begin{split}
 R(E)= & \frac{1}{\hbar ^2}(N_B(E)+1)\\
  \times & \frac{1}{N}\sum_{k}F_{11}(k)F_{11}^*(k)[\delta (E-E(k))+\delta(E+E(k))].
\end{split}
\end{equation}
After performing the summation over k in the continuum limit, introducing the
quadratic density of state of the phonons corresponding to the 3D case,  the phonon spectral density is :
\begin{equation}
  \label{}
 R(E)= R_oE^3g(E),
\end{equation}
with
\begin{equation}
 R_o=\frac{(\sigma_h)^2}{8\pi^2\hbar \rho c_l^5}.
\end{equation}
The quantity $R_o$ contains all material parameters. The cubic dependence is due
to the quadratic phonon density of state, and the function g(E) is a function
of the energy and of the geometrical parameters of the QD \citep{calculdensitespectralcomplet}:
\begin{equation}
\begin{split}
g(E)= \int_{-\frac{\pi}{2}}^{\frac{\pi}{2}}\zeta \cos \zeta \exp \left[-\frac{(l_{\bot})^2E^2}{2\hbar ^2c_l^2} \right. \\
 \times \left. \left(cos^2\zeta +\frac{l_z^2}{(l_{\bot})^2}sin^2\zeta \right)\right]d\zeta,
 \end{split}
\end{equation}
where $ \zeta $ is the angle of the wave vector k with respect to the normal to the z
direction.

Describing coupling of the hole to phonons confined in a nanowire the same way
as in a 3D semi conductor bulk matrix seems a rough approximation. So we also
considered the 1D case in which the nanowire is taken as an infinitely thin monomode wave
guide. The phonon density of state is constant and we consider that only the
phonons propagating along the nanowire (z direction) can couple to the hole
whose wavefunction is taken as $\psi_h(r)\propto\exp[-(1/2)(z/l_z)^2]$.
The spectral density becomes:
\begin{equation}
  \label{}
R_{1D}(E)\propto Eg_{1D}(E),
\end{equation}
with
\begin{equation}
  \label{}
 g_{1D}(E)= \exp \left[-\frac{l_z^2E^2}{2\hbar ^2c_l^2}\right].
\end{equation}
The phonon spectral density is linear with energy but the geometrical factor
$g_{1D}(E)$ has the same gaussian energy dependence as g(E).

\begin{figure}
\resizebox{0.56\textwidth}{!}{\includegraphics[width=0.4\textwidth]{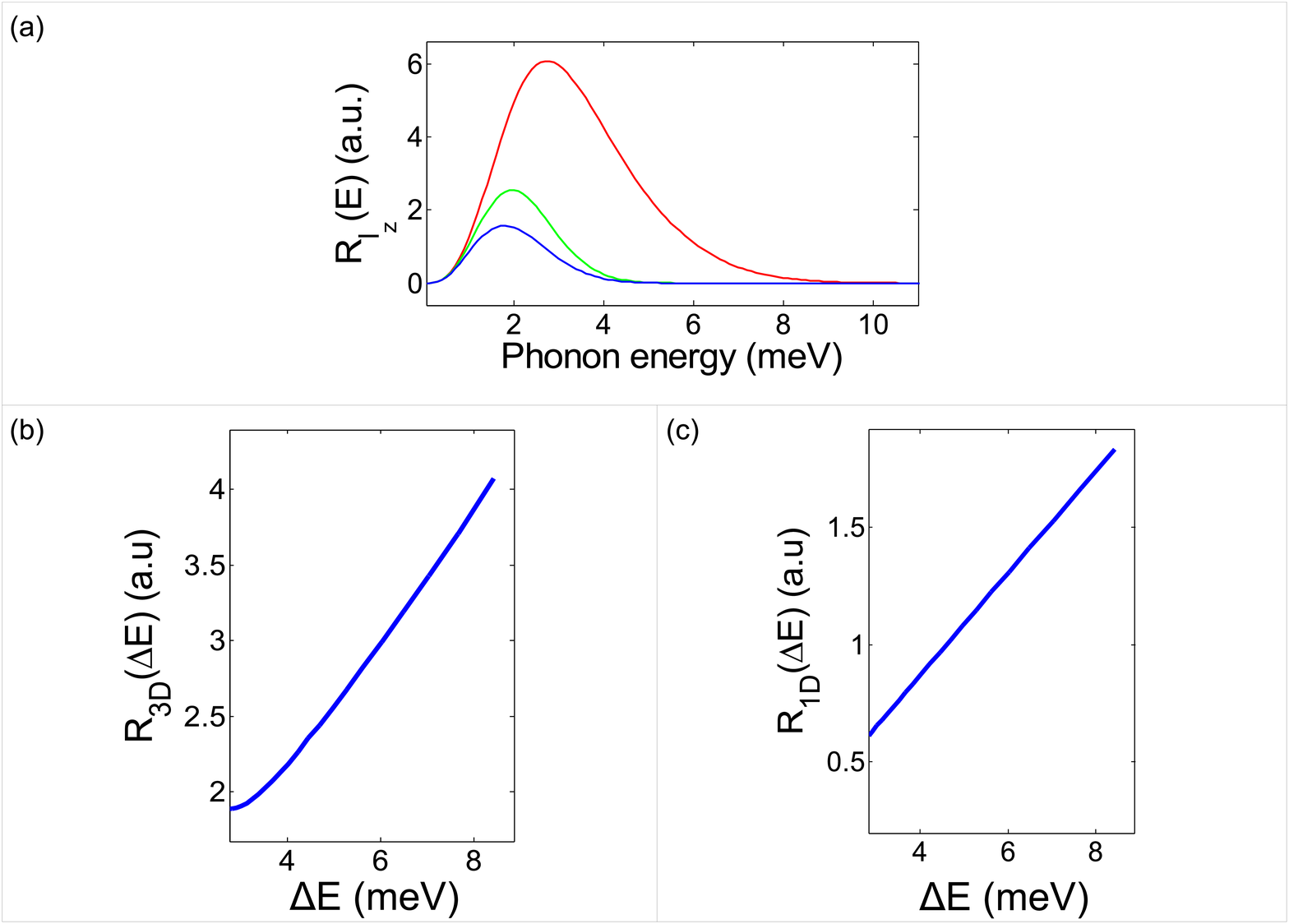}}
\caption{(Color online) (a) 3D Phonon spectral density for 3 different QD sizes along the z
direction (red curve: $l_z$ =2 nm, green curve: $l_z$=4 nm, blue curve: $l_z$=6 nm). Smaller QD z dimension shifts the maximum of the coupling constant
toward the higher energies, and enhances its relative value. (b) Calculated 3D
hole-phonon coupling efficiency vs transition energy $\Delta E$. (c) Calculated 1D
hole-phonon coupling efficiency vs transition energy $\Delta E$.} \label{fig:R3tailles}
\end{figure}

\begin{figure}
\resizebox{0.4\textwidth}{!}{\includegraphics*[0,50][800,700]{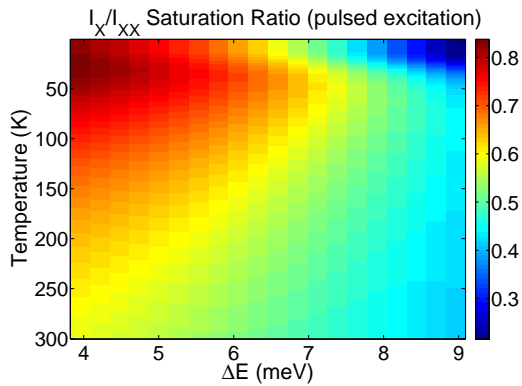}}
\caption{(Color online) Calculated saturation ratio Ix/Ixx under pulsed excitation as a
function of $\Delta E$ and temperature. The color legend represents Ix/Ixx
saturation ratio. The cold colors show low exciton intensity and the hot color
high intensity.} \label{fig:rapport}
\end{figure}

\begin{figure}
\resizebox{0.45\textwidth}{!}{\includegraphics*[0,150][800,700]{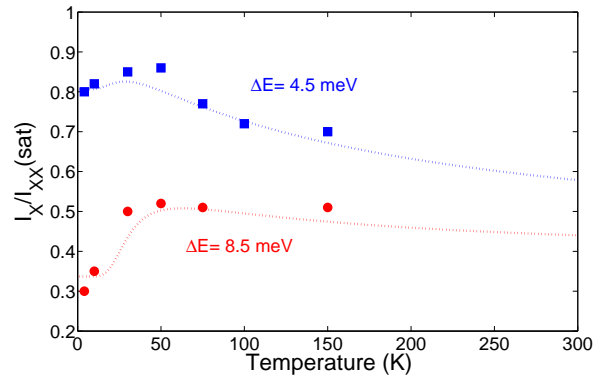}}
\caption{(Color online) Experimental exciton-biexciton saturation ratio under pulsed excitation versus
temperature for two QDs with  respectively $\Delta E=4.5$ meV (blue squares) and  $ \Delta E =8.5$ meV
(red circles). Dashed lines are vertical cut (fixed $ \Delta E $) from the model used in figure \ref{fig:rapport}.} \label{fig:ratiotemp}
\end{figure}

Since at $T=4$K, $\gamma_{sp1}(4K) \propto R(\Delta E)$, the BS to DS spin flip rate
evolution from dot to dot can be described by calculating the phonon spectral
density for each QD.
Three calculated 3D phonon spectral densities corresponding to three different
$l_z$ values ($l_z=2,4,6$ nm) are plotted in figure \ref{fig:R3tailles}(a).  Because of
its increase with energy (linear for 1D, cubic for 3D),
 $R(\Delta E)$ corresponding to QDs with
smaller $l_z$ have their maximum enhanced and shifted toward the higher
energies. As a consequence, in small QDs, high energy phonons couple
more efficiently with the hole. We propose to explain qualitatively the increase
of the transition rate observed in figure \ref{fig:gammaovssplit} by this size effect.

 The hole-phonon coupling efficiency, $R_{l_z}(\Delta E)$ depends on the QD size $l_z$ and on the energy splitting $\Delta E$. We
determine the relation between $l_z$ and $\Delta E$ by evaluating the short
range hole-electron exchange energy in the QD:
\begin{equation}
  \label{}
\Delta E= \Delta E^{3d}\frac{1}{|\varphi^{3d}(0)|^2}\int |\psi_e(r)|^2|\psi_h(r)|^2d^3r,
\end{equation}
where $\psi_{e/h}(r)$ are the electron/hole wave functions, $\Delta E^{3d}$ is the
exchange energy in the bulk material ZnCdSe \citep{Puls} ($\Delta E^{3d}=0.19$ meV, for
a Cd$_{0.5}$Zn$_{0.5}$Se composition of the QD as
 measured in high resolution TEM  experiments \citep{JAP}), and $|\varphi^{3d}(0)|^2=
1 / \pi a_B^{*3}$ with ${a_B^*}$ the Bohr radius of the free exciton.
To match the measured values of $\Delta E$ (from 4 meV to 9 meV), we set the
lateral confinement parameter $l_\bot$ at 4nm. With $l_z$ linked to a
corresponding $\Delta E$ we make $l_z$ vary from 2 nm to 6 nm and we calculate
$R_{l_z}(\Delta E)$.

 The results, for the 3D and 1D cases described above, are plotted in figure \ref{fig:R3tailles} (b) and (c).
  Because of the simplicity of the considered exciton wave function, the aim of the calculation is
  neither to fit the experimental data nor
to obtain a quantitative estimation of the transition rates. However, we can
show that the hole-phonon coupling efficiency is increasing with $\Delta E$ in
both cases, which is not a trivial result as the cut-off imposed by the
dimensions of the QD makes this efficiency vanish for higher energy phonons.
The effect of the confinement dimensions on the phonon spectral density can be
a good explanation for the trend measured by the experiment in figure
\ref{fig:gammaovssplit}. The two situations considered here are extreme cases
and we can expect a real nanowire geometry to impose an intermediate behavior
for the hole phonon coupling.

\section{Exciton luminescence intensity}
\label{exciton}

This increase of the exciton-phonon coupling efficiency with energy splitting
has some consequences on the temperature dependent exciton-biexciton saturation
ratio under pulsed excitation. This calculated ratio is represented in figure
\ref{fig:rapport} as a function of temperature and energy splitting $\Delta E$.
We used the level scheme presented in figure \ref{fig:spectre} and the values
of the transition rates given in section \ref{DeltaE}. The relation between
$\gamma_{sp1}(4K)$ and $\Delta E$ is taken from the function used as a guide
for the eyes in figure \ref{fig:gammaovssplit}. We also considered the increase
of the dark exciton non radiative recombination $\Gamma_{nr}(T)$ with
temperature following an Arrhenius  law $\Gamma_{nr}(T) \sim exp(-E_a/k_B T)$,
with an activation energy $E_a$. For the two QDs studied in fig. \ref{fig:ratiotemp}
   we have measured  $E_a=30\pm5$ meV and we have taken $E_a=30$ meV in the model used in figures \ref{fig:rapport} and \ref{fig:ratiotemp}. The
particularity here is that the DS lifetime at $T=4$ K ($1/\Gamma_{nr}= 5$ ns) is very
short compared to values reported in other systems (up to $1 \mu s$
\citep{lounistempdep}) and is of the same order of
magnitude as the exciton radiative lifetime when temperature is raised up to
only several tens of K.  A characteristic temperature behavior is shown in
figure \ref{fig:rapport}. The exciton intensity is very small at low
temperature (particularly in the high splitting region), increases
 as temperature is increased up to $T=50$ K, and finally
 decreases with higher temperatures. The effect of repopulation of the bright
exciton due to the DS to BS spin flip is compensated and overwhelmed by the
exciton population loss caused by the faster dark exciton  non radiative
recombination. As shown in figure \ref{fig:ratiotemp}, representing the
exciton-biexciton ratio versus temperature for 2 different QDs with different
measured $ \Delta E $ ($ 4.5$ meV  and $ 8.5$ meV ), this exciton line intensity decrease is more
sensitive for  large $ \Delta E $. However, at 300K, the bright exciton is
still less luminescent for large splittings. It can be noted that the model
fits rather well the experimental data.

\section{Conclusion}

In summary, the exciton-phonon coupling efficiency is highly influenced by the
QD size. This can explain the observed enhancement of the bright to
dark spin flip rate with increasing splitting energy. As a result, the saturation intensity of exciton transition
is a lot weaker than the biexciton one. This effect is all the more important as the QD exchange splitting $\Delta E$ is larger. The domination of the biexcitonic line over the excitonic line
is preserved at high temperature despite the bright state repopulation owing to
 the temperature induced loss of excitonic population through the non
radiative dark exciton recombination.
This exciton weakness could suggest to use the biexciton line for high temperature single photon sources. In that case, the contamination by the exciton line could be reduced when the lines are broadened at high temperature.

\section{Acknowledgements}

We acknowledge support from the French National Research Agency (ANR) through the Nanoscience and Nanotechnology Program (Project BONAFO ANR-08-NANO-031-01) that provided a research fellowship for MdH. MEJ acknowledges financial support from the Nanosciences Foundation "Nanosciences, aux limites de la nano\'{e}lectronique" (RTRA).


\begin{thebibliography}{10}


\bibitem{Duan01} X. Duan, Y. Huang, Y. Cui, J. Wang, and C. M. Lieber, Nature
    \textbf{409}, 66  (2001)

\bibitem{Lu}   W. Lu and C. M. Lieber, Nature Materials \textbf{6}, 841 (2007)

\bibitem{The03}
C. Thelander, T. Martensson, M. T. Björk, B. J. Ohlsson, M. W. Larsson, L. R.
Wallenberg, and L. Samuelson, Appl. Phys. Lett. \textbf{83}, 2052, (2003).

\bibitem{Konenkamp} R. Könenkamp, R.C. Word, and C. Schlegel
Appl. Phys. Lett. \textbf{85}, 6004 (2004)

\bibitem{Kim} H. M. Kim, Y. H. Cho, H. Lee, S. I. Kim, S. R. Ryu, D. Y. Kim, T. W.
    Kang, and K. S. Chung, Nano Lett. \textbf{4}, 1059 (2004)


\bibitem{Duan03} X. Duan, Y. Huang, R. Agarwal, and C. M. Lieber, Nature
    \textbf{421}, 241 (2003)

\bibitem{Hochbaum}
A. I. Hochbaum, R.Chen, R. D. Delgado, W. Liang, E. C. Garnett, M. Najarian, A.
Majumdar, and P. Yang, Nature \textbf{451}, 163  (2008)

\bibitem{Cui}
Y. Cui, Q. Wei, H. Park, and C. M. Lieber, Science \textbf{293}, 1289 (2001)

\bibitem{Gud02}
M. S. Gudiksen, L. Lauhon, J. Wang, D. C. Smith,  and  C. M. Lieber, Nature
\textbf{415}, 617 (2002)

\bibitem{Bjork} 
M. T. Björk, B. J. Ohlsson, T. Sass, A. I. Persson, C. Thelander, M. H.
Magnusson, K. Deppert, L. R. Wallenberg, and  L. Samuelson, Nanoletters \textbf{2},
87 (2002); Appl. Phys. Lett. \textbf{80}, 1058 (2002)

\bibitem{Yang}  C. Yang, Z. Zhong, and C. M. Lieber, Science \textbf{310},  1304
    (2005)




\bibitem{singleemitterimamoglu} P. Michler, A. Kiraz, C. Becher, W. V.
    Schoenfeld, P. M. Petroff, L. Zang, E. Hu, and A. Imamoglu, Science
    \textbf{290}, 2282 (2000).


\bibitem{Moreau} E. Moreau, I. Robert, J.-M. G\'{e}rard, I. Abram, L. Manin, and V.
    Thierry-Mieg,
  Appl. Phys. Lett. \textbf{79}, 2865 (2001).

\bibitem{Santori}  C. Santori, D. Fattal, J. Vu\v{c}kovi\'{c}, G. S. Solomon, and Y.
    Yamamoto,
   Nature \textbf{419}, 594 (2002).

 \bibitem{Chang}
  W. H. Chang, W. Y. Chen, H. S. Chang, T. P. Hsieh, J. I. Chyi, and T. M. Hsu,
  Phys. Rev. Lett. \textbf{96}, 117401 (2006).


  \bibitem{Claudon} J. Claudon, J. Bleuse, N. S. Malik, M. Bazin, P. Jaffrennou, N. Gregersen, C. Sauvan, P. Lalanne, and J.-M. G\'{e}rard, Nature Photonics \textbf{4}, 174 (2010)

\bibitem{borgstrom}	M. T. Borgstr\"{o}m, V. Zwiller, E. M\"{u}ller, and  A. Imamoglu, Nano
    Lett. \textbf{5}, 1439 (2005)

\bibitem{TRIBU220k}
A. Tribu, G. Sallen, T. Aichele, R. Andr\'{e}, C. Bougerol, S. Tatarenko, J.-Ph. Poizat, and K. Kheng, Nanoletters \textbf{349}, 225 (1991).

\bibitem{dark}
V. D. Kulakovskii, G. Bacher, R. Weigand, T. K\"{u}mmell, A. Forchel, E. Borovitskaya, K. Leonardi and D. Hommel, Phys. Rev. Lett. \textbf{82}, 1780 (1999);
L. Besombes, K. Kheng, and D. Martrou, Phys. Rev. Lett.  \textbf{85}, 425 (2000).



\bibitem{sallen}
G. Sallen, A. Tribu, T. Aichele, R. Andr\'{e}, L. Besombes, C. Bougerol, S.
Tatarenko, K. Kheng, and J.-Ph. Poizat, Phys. Rev. B \textbf{80}, 085310
(2009).

\bibitem{quantumwelldarkexc}
Y. Chen, B. Gil, P. Lefebvre, and H. Mathieu, Phys. Rev. B \textbf{37}, 6429
(1988).

\bibitem{quantumwire}
Y. Chen, Phys. Rev. B \textbf{41}, 10604 (1990).

\bibitem{efrosexp}
M. Nirmal, D. J. Norris, M. Kuno,  M. G. Bawendi, Al. L. Efros, and M. Rosen,
Phys. Rev. Lett. \textbf{75}, 3728 (1995).

\bibitem{influencedsonqd}
M. Reischle, G. J. Beirne, R. Rossbach, M. Jetter, and P. Michler, Phys. Rev.
Lett. \textbf{101}, 146402 (2008).

\bibitem{lounistempdep}
O. Labeau, P. Tamarat, and B. Lounis, Phys. Rev. Lett. \textbf{90}, 257404 (2003).


\bibitem{holespinflip}
L. M. Woods, T. L. Reinecke, and R. Kotlyar, Phys. Rev. B \textbf{69}, 125330
(2004)


\bibitem{phonon-exciton_spectraldensity}
A. Grodecka, L. Jacak, P. Machnikowski, and K. Roszak,  arXiv:cond-mat/0404364
(2004).

\bibitem{Hertog}
M. Den Hertog, M. Elouneg-Jamroz, E. Bellet-Amalric, S. Bounouar, C. Bougerol, R. André, Y. Genuist, J.-Ph. Poizat, K. Kheng and S. Tatarenko, J. Crystal Growth
 \textbf{323},   330 (2011)

\bibitem{JAP} M. Den Hertog, M. Elouneg-Jamroz, E. Bellet-Amalric, S. Bounouar,
    C. Bougerol, R. André, Y. Genuist, J.-Ph. Poizat, K. Kheng and S. Tatarenko,
 J. Appl. Phys. \textbf{110}, 034318 (2011).

\bibitem{lucienphononbroad}
L. Besombes, K. Kheng, L. Marsal, and H. Mariette, Phys. Rev. B \textbf{63},
155307 (2001).

\bibitem{Takagahra}
T. Takagahara, Phys. Rev. B \textbf{60}, 2638 (1999)

\bibitem{efros}
Al. L. Efros, and M. Rosen, 
M. Kuno, M. Nirmal, D. J. Norris, and M. Bawendi,
Phys. Rev. B \textbf{54}, 4843 (1996).



\bibitem{calculdensitespectralcomplet}
A. Grodecka, C. Weber, P. Machnikowski, and A. Knorr, Phys. Rev. B \textbf{76},
205305 (2007)

\bibitem{Puls}
J. Puls, F. Henneberger, M. Rabe, and A. Siarkos,  Journal of Crystal Growth
\textbf{184/185}, 787 (1998)






\end{thebibliography}
\end{document}